\documentstyle[12pt]{article}

\def\be{\begin{equation}}
\def\ee{\end{equation}}

\def\H{{\cal H}}

\def\Q{{\cal Q}}
\def\R{\!^{(3)}{\cal R}}
\def\GE{\raise.8ex\hbox{$>$\kern-1.8ex\lower1.25ex\hbox{$\approx$}}}
\def\QHJ{quantum Hamilton--Jacobi}
\title{PURE QUANTUM SOLUTIONS OF BOHMIAN QUANTUM GRAVITY}
\author{FATIMAH SHOJAI$^{1,3}$\thanks{Email: FATIMAH@THEORY.IPM.AC.IR}
and ALI SHOJAI$^{2,3}$\thanks{Email: SHOJAI@THEORY.IPM.AC.IR} \\
$^1$ Physics Department, Iran University of Science and
Technology,\\ P.O.Box 16765--163, Narmak, Tehran, IRAN\\ $^2$
Physics Department, Tarbiat Modares University,\\ P.O.Box
14155--4838, Tehran, IRAN\\ $^3$ Institute for Studies in
Theoretical Physics and Mathematics,\\ P.O.Box 19395-5531, Tehran,
IRAN}
\date{}
\begin{document}
\maketitle \vspace{-1cm}
\begin{abstract}
In this paper we have investigated the {\it pure quantum\/}
solutions of Bohmian quantum gravity. By pure quantum solution we
mean a solution in which the quantum potential cannot be ignored
with respect to the classical potential, especially in Bohmian
quantum gravity we are interested in the case where these two
potentials are equal in their magnitude and in fact their sum is
zero. Such a solutions are obtained both using the perturbation
and using the linear field approximation.

\end{abstract}
\section{INTRODUCTION}
The idea of quantization of the gravitational field is supported
by different reasons. Quantum gravity is a good candidate for a
fundamental theory because it is expected to avoid the general
relativity singularities. This theory must lead us to some
understanding of the initial conditions of the universe, inflation
epoch, and so on. Furthermore, if one looks for the unification of
gravity with the other fields, the fundamental description of
gravity should be achieved by its quantization, just like the
other fields.

Many approaches exist for quantization of gravity. One of them is
the canonical quantization. Depending on the choice of the
canonical variables, it can be divided into many classes. One of
them is the geometrodynamical approach, which uses ADM
decomposition. In this decomposition, dynamical variables are the
three dimensional metric ($h_{ij}(x)$) having six independent
elements. A specific character of quantum gravity is the existence
of constraint, resulted from the four dimensional diffeomorphism
invariance. Quantum gravity hamiltonian is given by:
\be
\widehat{H}=\int d^3x
\left(N^{\bot}\widehat{\H}^{\bot}+N^i\widehat{\H}_i\right) \ee
where $N^{\bot}$ and $N^i$ are the lapse and shift functions
respectively. These are the remaining non--dynamical four elements
of the space--time metric. $\widehat{\H}^{\bot}$ and
$\widehat{\H}_i$ are the time and three diffeomorphism invariance
constraints. According to the Dirac's canonical quantization
scheme, the constraints should kill the wavefunction of the
universe. In terms of the canonical operators, we have the WDW
equation:
\be
\left [h^{-q}\frac{\delta}{\delta
h_{ij}}h^qG_{ijkl}\frac{\delta}{\delta h_{kl}}+\sqrt{h}\R
+\frac{1}{2\sqrt{h}}\frac{\delta^2}{\delta\phi^2}-\frac{1}{2}
\sqrt{h}h^{ij}\partial_i\phi\partial_j\phi-\frac{1}{2}\sqrt{h}V(\phi)\right
]\Psi=0 \label{WDWE}
\ee
and the momentum constraint:
\be
i\left [2\nabla_j\frac{\delta}{\delta
h_{ij}}-h^{ij}\partial_j\phi\frac{\delta}{\delta\phi}\right
]\Psi=0 \label{MOME} \ee where $\phi$ denotes the matter field,
$G_{ijkl}$ is the DeWitt metric with signature $(-+++++)$, $q$ is
the ordering parameter, and $h=\det (h_{ij})$.

Although in this manner, the canonical quantization is applied to
gravity, but in general the application of the standard quantum
mechanics to gravity leads to many difficulties. Some of them are
conceptual and some others are related to the mathematical
structure of quantum mechanics. Because of the above problems,
some physicists are interested in alternatives to quantum
mechanics. Among these alternatives, deBroglie--Bohm
theory\cite{Bohm} is of special importance. This theory has some
special characteristics, such as:
\begin{itemize}
  \item It provides an explanation of the individual events. In
  addition, its statistical content is just like the standard
  quantum mechanics.
  \item It presents an objectively real point of view of nature.
  The system exists independent of whether we measure it or not.
  So, it is not merely an abstract description of the system.
  \item The outcome of any measurement process is described by a
  causally related series of individual processes.
  \item All the particles have a well defined trajectory in the
  space--time, which can be evaluated from the quantum
  Hamilton--Jacobi function and the initial conditions.
  \item The wavefunction plays two roles. Its norm leads to the
  probability distribution, whereas its phase is proportional to
  the quantum Hamilton--Jacobi function.
  \item The guidance equation in this theory leads to the time
  evolution of dynamical variables independent of whether the
  wavefunction depends or not depends on time.
  \item deBroglie--Bohm theory is one of the appropriate theories
  of the universe, because it can be applied to the world as a
  whole, without any division to observer and observant.
\end{itemize}

In the deBroglie--Bohm theory the quantum Hamilton--Jacobi
function satisfies some Hamilton--Jacobi equation modified by the
quantum potential:
\be
\frac{\partial S}{\partial t}+\frac{|\vec{\nabla}S|^2}{2m}+V+\Q=0
\label{QHJE} \ee where the quantum potential is given by:
\be
\Q=-\frac{\hbar^2}{2m}\frac{\nabla^2\sqrt{\rho}}{\sqrt{\rho}} \ee
and $S$ is the Hamilton--Jacobi function, and $\rho$ is the
ensemble density. Furthermore, in order to preserve the
probability conservation, the continuity equation should be
satisfied:
\be
\frac{\partial\rho}{\partial t}+\vec{\nabla}\cdot\left (
\rho\frac{\vec{\nabla}S}{m}\right )=0 \label{CONE} \ee

The equations (\ref{QHJE}) and (\ref{CONE}) can be composed in
such a way that the Schr\"{o}dinger equation is resulted (setting
$\Psi=\sqrt{\rho}\exp (iS/\hbar)$). But it must be noted that the
physical content of the deBroglie--Bohm and the standard quantum
mechanics are completely different. This can be seen easily from
some of the above mentioned characteristics of the deBroglie--Bohm
theory. In the other words, the polar decomposition of the
wavefunction is the simplest way of obtaining the quantum
Hamilton--Jacobi and continuity equations. But the fact that one
must include some quantum potential with that specific form, can
be understood by physical intuition. In ref.\cite{ATHESIS} some
arguments are presented that leads one to write the \QHJ\ equation
and the continuity eqation without any reference to the
Schr\"odinger equation.

Note that in the \QHJ\ equation, setting $\Q=0$ one gets the
classical equations of motion. This means that the {\it classical
limit\/} can be defined as the case in which both the quantum
potential and the quantum force may be ignored with respect to the
classical counterparts, i.e. $|\Q|\ll |V|$ and
$|-\vec{\nabla}\Q|\ll|-\vec{\nabla}V|$. On the other hand one can
define an opposite limit which can be called {\it the pure quantum
limit\/} where neither quantum potential nor quantum force can be
ignored with respect to the classical counterparts, i.e.
$|\Q|\GE|V|$ and $|-\vec{\nabla}\Q|\GE|-\vec{\nabla}V|$. In the
classical limit the classical trajectory is slightly modified by
quantum correction, while in the pure quantum limit one gets a
trajectory that is not similar to any classical solution. Here we
are interested in the case that the sum of the quantum and
classical potentials are equal to the total energy of the system,
i.e. $-\partial S/\partial t=V+\Q$. Thus from the \QHJ\ equation,
$\vec{\nabla}S=0$ and the system is at rest. So this is a stable
situation of the system. It must be noted here that vanishing
phase gradient is an important case in the non--relativistic
de-Broglie--Bohm theory. For example, this occurs in the ground
state of hydrogen atom or all the energy eigenstates of one
dimensional harmonic oscillator.

By expressing the wavefunction of the universe in the polar form,
the equations (\ref{WDWE}) and (\ref{MOME}) lead to: \be
G_{ijkl}\frac{\delta S}{\delta h_{ij}}\frac{\delta S}{\delta
h_{kl}}+\frac{1}{2\sqrt{h}}\left (\frac{\delta
S}{\delta\phi}\right )^2-\sqrt{h}\left (\R-\Q_G\right
)+\frac{\sqrt{h}}{2}h^{ij}\partial_i\phi\partial_j\phi
+\frac{\sqrt{h}}{2}\left (V(\phi)+\Q_M\right )=0 \label{abcd}\ee
where the gravity quantum potential is given by: \be
\Q_G=-\frac{1}{\sqrt{h}\sqrt{\rho}}\left
(G_{ijkl}\frac{\delta^2\sqrt{\rho}}{\delta h_{ij}\delta
h_{kl}}+h^{-q}\frac{\delta h^qG_{ijkl}}{\delta h_{ij}}\frac{\delta
\sqrt{\rho}}{\delta h_{kl}}\right ) \ee and the matter quantum
potential is defined as: \be
\Q_M=-\frac{1}{h\sqrt{\rho}}\frac{\delta^2\sqrt{\rho}}{\delta\phi^2}
\ee and the continuity equation: \be \frac{\delta}{\delta
h_{ij}}\left [2h^qG_{ijkl}\frac{\delta S}{\delta
h_{kl}}\rho\right ]+\frac{\delta}{\delta\phi}\left [
\frac{h^q}{\sqrt{h}}\frac{\delta S}{\delta\phi}\rho\right ]=0 \ee
The momentum constraint leads to: \be
2\nabla_j\frac{\delta\sqrt{\rho}}{\delta
h_{ij}}-h^{ij}\partial_j\phi\frac{\delta\sqrt{\rho}}{\delta\phi}=0;
\ \ \ \ 2\nabla_j\frac{\delta S}{\delta
h_{ij}}-h^{ij}\partial_j\phi\frac{\delta S}{\delta\phi}=0 \ee The
guiding equations are: \be \frac{\delta S}{\delta
h_{kl}}=\pi^{kl}\equiv\sqrt{h}\left (K^{kl}-h^{kl}K\right );\ \ \
\ \ \frac{\delta
S}{\delta\phi}=\pi_\phi\equiv\frac{\sqrt{h}}{N^\bot}
\dot{\phi}-\sqrt{h}\frac{N^i}{N^\bot}\partial_i\phi \ee where
$K^{ij}$ is the extrinsic curvature. Since in the WDW equation,
the wavefunction is in the ground state with zero energy, then the
stability condition of the metric and of the matter field is: \be
h^{ij}\partial_i\phi\partial_j\phi +V(\phi)-2\R+\Q_M+2\Q_G=0 \ee
which is a pure quantum solution. This equation can simply
derived from the equation (\ref{abcd}) by setting all functional
derivatives of $S$ equal to zero. In this paper we first solve the
pure quantum case perturbatively and then investigate the pure
quantum solution without matter in the linear field approximation.
\vspace{-1cm}
\section{Perturbative Solution of Pure Quantum Cases}
First we shall discuss the pure quantum solutions using the method
of perturbation in solving the Bohmian quantum gravity equations.
Recently\cite{cper} a perturbative method of solving the classical
Hamilton--Jacobi equations of general relativity has been
presented. In ref. \cite{per}, this approach is adopted to solving
Bohmian quantum gravity equations. In this method the \QHJ\
function and the norm of the wavefunction have been expanded in
terms of powers of spatial gradiants of the metric and matter
fields. (the so called long wavelength approximation) Solutions up
to the second order are derived. At each order the form of $S$ and
$\sqrt{\rho}$ functionals is chosen in such a way that they are 3
diffeomorphism invariants. According to \cite{per}, we set: \be
\sqrt{\rho}=\exp(\Omega);\ \ \ \ \ \ \Omega=\sum_{n=0}^\infty
\Omega^{(2n)} \ee In the pure quantum case which we are interested
in here, the continuity equation will be satisfied trivially.
Therefore at each order it is sufficient to solve \QHJ\  equation.
So we must set: \be -\sqrt{h}\left (\R-\Q_G\right
)+\frac{1}{2}\sqrt{h}\partial_i\phi\partial^i\phi+\frac{1}{2}\sqrt{h}\left
(V+\Q_M\right )=0 \label{xyz}\ee \par
 \vspace{-1cm}
\subsection{zeroth order solution}
In this order  the \QHJ\  equation is:
\be
-2\sqrt{h}G_{ijkl}\left (\frac{\delta^2\Omega^{(0)}}{\delta
h_{ij}\delta h_{kl}}+\frac{\delta\Omega^{(0)}}{\delta
h_{ij}}\frac{\delta\Omega^{(0)}}{\delta h_{kl}}\right )+
(q+3/2)h_{ij}\frac{\delta\Omega^{(0)}}{\delta
h_{ij}}-\frac{\delta^2\Omega^{(0)}}{\delta\phi^2}-\left
(\frac{\delta\Omega^{(0)}}{\delta\phi}\right )^2+hV(\phi)=0
\label{haho} \ee The appropriate choice of $\Omega^{(0)}$ which is
3 diffeomorphism invariant is:
\be
\Omega^{(0)}=\int d^3x\ \sqrt{h}J(\phi) \ee Substituting this in
the relation (\ref{haho}), we have:
\be
J''-\frac{3}{2}(q+5)J-\sqrt{h}\left (\frac{3}{4}J^2-J'^2\right
)-\sqrt{h}V(\phi)=0 \ee Since the terms containing metric must
cancel each other, we have the two following equations:
\be
J''-\frac{3}{2}(q+5)J=0;\ \ \ \ \ \ \frac{3}{4}J^2-J'^2+V(\phi)=0
\ee These two equations can be easily solved to obtain the $K$
function and fix the form of $V$. We choose the solution as:
\be
J=J_0\exp(\alpha\phi);\ \ \ \ \ \alpha^2=\frac{3}{2}(q+5) \ee so
that:
\be
V(\phi)=\left (\alpha^2-\frac{3}{4}\right ) J_0^2\exp
(2\alpha\phi) \ee Therefore we have:
\be
\Omega^{(0)}=\int d^3x \sqrt{h}J_0\exp(\alpha\phi) \ee
\par
\vspace{-1cm}
\subsection{second order solution}
In the second order, the \QHJ\  equation for the pure quantum case
is:
\be
-\sqrt{h} G_{ijkl}\left (\frac{\delta^2\Omega^{(2)}}{\delta
h_{ij}\delta h_{kl}}+2\frac{\delta\Omega^{(0)}}{\delta h_{ij}}
\frac{\delta\Omega^{(2)}}{\delta h_{kl}}\right )
+\frac{q+3}{2}h_{ij}\frac{\delta\Omega^{(2)}}{\delta
h_{ij}}-\frac{1}{2}\frac{\delta^2\Omega^{(2)}}{\delta\phi^2}-
\frac{\delta\Omega^{(0)}}{\delta\phi}\frac{\delta\Omega^{(2)}}{\delta\phi}
-2h\left (\R-\frac{1}{2}\nabla_i\phi\nabla^i\phi\right )=0 \ee On
using the zeroth order equation and considering terms with and
without metric separately, one arrives at:
\be
-\sqrt{h}G_{ijkl}\frac{\delta^2\Omega^{(2)}}{\delta h_{ij}\delta
h_{kl}}+\frac{q+3}{2}h_{ij}\frac{\delta\Omega^{(2)}}{\delta
h_{ij}}-\frac{1}{2}\frac{\delta^2\Omega^{(2)}}{\delta\phi^2}=0 \ee
\be
Jh_{kl}\frac{\delta\Omega^{(2)}}{\delta
h_{kl}}-2J'\frac{\delta\Omega^{(2)}}{\delta\phi}-4\sqrt{h}\left
(\R-\frac{1}{2}|\nabla\phi|^2\right )=0 \ee In order to solve
these equations we use the conformal transformation presented in
\cite{cper,per}. We set $h_{ij}=F^2(\phi)f_{ij}$ and assuming \be
\Omega^{(2)}=\int d^3x \sqrt{f}\left [
\widetilde{\R}L(\phi)+f^{ij}M(\phi)\partial_i\phi\partial_j\phi\right
] \ee where a $\sim$ over any quantity means that it is calculated
with respect to the $f_{ij}$ metric, and
\be
\frac{\delta\Omega^{(2)}}{\delta\phi}=A[F(\phi)]
\frac{\delta\Omega^{(2)}}{\delta
F} \ee one gets the following equations:
\be
AF'=\frac{F^2}{12};\ \ \ \ \left (\frac{19}{12}-\frac{q}{2}\right
)F=F'\frac{dA}{dF};\ \ \ \ \left ( \frac{JF}{2A}-2J'\right
)L'-4F=0\ee \be -\left ( \frac{JF}{2A}-2J'\right
)M'-16F''+8\frac{F'^2}{F}+2F=0;\ \ \ -\left (
\frac{JF}{2A}-2J'\right )M-8F'=0\ee with the solution: \be q=3;\ \
\ \alpha^2=12;\ \ \ A=\frac{\alpha}{3}F;\ \ \ F=F_0
e^{\phi/4\alpha}\ee \be
M=-\frac{2F_0}{J_0(3/2-2\alpha^2)}e^{-(\alpha-1/4\alpha)\phi};\ \
\
L=-\frac{4F_0}{J_0(2\alpha^2+3/8\alpha^2-2)}e^{-(\alpha-1/4\alpha)\phi}
\ee writing $\Omega^{(2)}$ in terms of the original metric, we
have:
\be
\Omega^{(2)}=\int d^3x
\sqrt{h}\frac{e^{-\alpha\phi}}{J_0(2\alpha^2+3/8\alpha^2-2)}\left
[ \R+\frac{12\alpha^2-11}{6-8\alpha^2}|\nabla\phi|^2\right ] \ee

Up to this point, we have found the stable solutions of Bohmian
quantum gravity up to the second order approximation. This method
can be applied to higher order approximations, too.

As we have presented some arguments in \cite{per}, in
deBroglie--Bohm theory, the classical limit and the long
wavelength approximation are two different domains, in general.
This is resulted from the nature of the quantum potential. So we
have the freedom of choosing any arbitrary value of quantum
potential in the long wavelength approximation. For example here
the sum of the quantum and classical potentials is equal to zero.
\vspace{-1cm}
\section{Pure quantum solutions without matter}
In the previous section dependence of the wavefunction on the
metric and the matter field is derived for the stable case
perturbatively. Here we attack the problem with another strategy.
For more simplicity we neglect the matter field here, and accept
the following form for $\Omega$:
\be
\Omega=\int d^3x\sqrt{h}\R \ee For simplicity we have chosen
$\Omega$ containing second spatial gradiants of metric. We want to
find the spatial metric for the pure quantum state of the
universe. Setting $q=-3/2$ (which sets the second term in $\Q_G$
equal to zero, and thus simplifies the equations) and $\delta
S/\delta h_{ij}=0$ in equation (\ref{xyz}), and ignoring the
matter field we have:
\be
-\sqrt{h}\R-\hbar^2G_{ijkl}\frac{\delta^2\Omega}{\delta
h_{ij}\delta h_{kl}}-\hbar^2G_{ijkl}\frac{\delta\Omega}{\delta
h_{ij}}\frac{\delta\Omega}{\delta h_{kl}}=0 \ee Substituting
$\Omega$ in the above relation, we have the following equation for
the scalar curvature:
\be
\hbar^2\left ( \frac{\R}{8}+\frac{\R^2}{4}+
\frac{\nabla^2\R}{2}-\R_{ij}\R^{ij}-\nabla_i\nabla_j\R^{ij}\right
) -\sqrt{h}\left [ \R+\hbar^2\left
(\R_{ij}\R^{ij}-\frac{\R^2}{8}\right )\right ]=0 \label{UU} \ee
This equation is a second order relation of the scalar curvature
and Ricci tensor, including their spatial gradiants. For finding
the solution we use linear approximation. Suppose the spatial
metric be of the form $h_{ij}=\eta_{ij}+\epsilon_{ij}$. Up to
first order in $\epsilon$ the equation (\ref{UU}) gives
\be
\partial^2\epsilon-\partial_i\partial_j\epsilon^{ij}=0
\label{ii}\ee In order to solve the above equation, one can expand
$\epsilon^{ij}$ in terms of the spherical harmonics:
\be
\epsilon^{ij}=\sum_{l,m}\alpha^{ij}_{lm}f_l(r)Y_{lm}(\theta,\varphi)
\ee and then substitute this in the equation (\ref{ii}). Finally
we find:
\be
\alpha^{ij}_{lm}=\left (
\begin{tabular}{ccc}$\alpha_{lm}$&0&0\\0&0&0\\0&0&0\end{tabular}\right )
;\ \ \ \ \ \ f_l(r)=r^{-l(l+1)/2} \ee Therefore the space--time
metric is:
\be
g_{\mu\nu}=\left (
\begin{tabular}{cccc}1&0&0&0\\
0&$1+\sum_{lm}\alpha_{lm}r^{-l(l+1)/2}Y_{lm}(\theta,\varphi)$&0&0\\
0&0&$r^2$&0\\0&0&0&$r^2\sin^2\theta$\end{tabular}\right ) \ee
where we have chosen $N^\bot=1$ and $N_i=0$. Some mathematical
manipulations result that $\partial_t$ is a killing vector leading
to the stationarity of the space--time metric. Also
$\partial_\varphi$ is another killing vector provided $m=0$. Since
the parameter $t$ is not appeared in the space--time metric, our
coordinate frame is consistent with the killing vector
$\partial_t$. In addition because $N^\bot=1$ and $N_i=0$, the
components of the Ricci tensor are:
\be
{\cal R}_{00}=0;\ \ \ \ {\cal R}_{0i}=0;\ \ \ \  {\cal
R}_{ij}=\R_{ij} \ee This means ${\cal R}=\R$. Using the above
relations one can see that not all the components of the Ricci
tensor are zero. For example ${\cal R}_{22}\neq 0$. Therefore the
quantum effects, change the classical solution (flat space--time)
to a curved space--time. This fact is shown in a completely
different way previously\cite{GEO}. \vspace{-1cm}
\section{conclusion}
Pure quantum solutions, i.e. the solutions in which the classical
potential and the quantum potential are of the same order are
important, because they represent cases which are not in any way
similar to the classical solutions. They have not classical limit.

Pure quantum solutions are very important in the non-relativistic
de-Broglie--Bohm theory. Many of stationary states, like the
ground state of hydrogen atom, or all the energy eigenstates of
one dimensional harmonic oscillator are pure quantum states in
which system is at rest. Such a solution may be prepared in the
laboratory as a state very different from classical mechanics.

For the quantum Cosmology case there is a difference. Since the
universe is prepared in some specific state and cannot be
reprepared in another way by us. The universe is believed thave
classical limit and thus cannot be a pure quantum state. But it
may be argued that at some times (times very close to BigBang,
say) the state might be pure quantum state.

In the case of quantum gravity, we can prepare systems in pure
quantum states at least in principle. So the investigation of
such states is not only a mathematical exercise in the
de-Broglie--Bohm quantum gravity and Cosmology, but also it may
be useful for (i) the initial times of the universe, (ii) some
specially prepared gravitational systems.

In this paper we studied the problem of how to solve the Bohmian
quantum gravity equations in the case of pure quantum limit. We
saw that there is at least two ways. First, one can use the
method of perturbation and the second way is not perturbative but
is written in the linear field approximation. 
\end{document}